\documentclass[
  aps,
  prl,
  reprint,
  floatfix,
  nofootinbib,
  preprintnumbers,
  superscriptaddress,
  nolongbibliography
]{revtex4-2}

\usepackage{amsmath,amssymb,graphicx,mathtools,hyperref, color}
\hypersetup{colorlinks=true,linkcolor=blue,citecolor=blue,urlcolor=blue}

\allowdisplaybreaks[3]

\newcommand{\M}{\mathcal{M}}
\newcommand{\Ima}{\operatorname{Im}}
\newcommand{\Rea}{\operatorname{Re}}
\newcommand{\dd}{\mathrm{d}}
\newcommand{\Ls}{\Lambda}
\newcommand{\GeV}{\mathrm{GeV}}
\newcommand{\cm}{\mathrm{cm}}

\begin{document}

\title{S-matrix bootstrap bounds on self-interacting dark matter}
\preprint{USTC-ICTS/PCFT-26-43}

\author{Qing Chen}
\email{qingchen@aust.edu.cn}
\affiliation{Center for Fundamental Physics, School of Mechanics and Physics, Anhui University of Science and Technology, Huainan, Anhui 232001, China}

\author{Zhuo-Hui Wang}
\email{wzh33@mail.ustc.edu.cn}
\affiliation{Interdisciplinary Center for Theoretical Study, University of Science and Technology of China, Hefei, Anhui 230026, China}

\author{Shuang-Yong Zhou}
\email{zhoushy@ustc.edu.cn}
\affiliation{Interdisciplinary Center for Theoretical Study, University of Science and Technology of China, Hefei, Anhui 230026, China}
\affiliation{Peng Huanwu Center for Fundamental Theory, Hefei, Anhui 230026, China}

\begin{abstract}
Self-interacting dark matter turns the structure of galactic halos into a direct requirement on a low-energy scattering amplitude. We show that, for weakly coupled scalar dark matter, this requirement implies a much stronger mass bound on the dark matter particle than partial-wave unitarity alone. Using analyticity, crossing symmetry, locality and partial-wave unitarity, we compute the maximal allowed threshold amplitude with a dispersive primal S-matrix bootstrap, assuming only a weakly coupled EFT below a scale $\Ls$ and allowing arbitrary UV particle content above $\Ls$. For the benchmark self-interaction cross section $\sigma_{\rm self}=10^{-24}(M/\GeV)\cm^2$, the mass of a generic weakly coupled scalar satisfies $M\lesssim 0.3\,\GeV$ in the controlled EFT regime. If dark matter is a derivative-dominated pseudo-Nambu-Goldstone boson, the mass bound is lowered to the MeV scale or below, depending on the hierarchy $M/\Lambda$.
\end{abstract}

\maketitle

The particle nature of dark matter remains one of the central open questions at the interface of particle physics, cosmology, and astrophysics. Cosmological observations have firmly established its abundance~\cite{Planck:2018vyg}, but they do not yet reveal its mass or its non-gravitational interactions. Galactic structure offers a rare additional clue. The inferred inner profiles and substructure of halos show persistent tensions with collisionless cold dark matter simulations, including the core-cusp problem and related small-scale puzzles~\cite{Moore:1994yx,Flores:1994gz,Navarro:1996gj,Navarro:2008kc,deBlok:2009sp,Bullock:2017xww,Sales:2022ich}. A compelling possibility is that dark matter has appreciable self-interactions~\cite{Spergel:1999mh,Adhikari:2022sbh,Nadler:2023nrd,Fischer:2025rky,Yu:2025tmp}. A self-interacting cross section per unit dark-matter mass in the range $\sigma_{\rm self}/M\sim0.1-10\,\cm^2/{\rm g}$ can modify the internal structure of galactic halos while preserving the success of cold dark matter on large scales~\cite{Tulin:2013teo,Kaplinghat:2015aga,Tulin:2017ara}.

Self-interacting dark matter (SIDM) links halo phenomenology directly to low-energy particle scattering. If the cross section per unit mass is fixed by astrophysical considerations, a heavier dark matter particle requires a larger nonrelativistic scattering amplitude. The question is therefore not only whether self-interactions are allowed, but how large the underlying amplitude can be in a consistent quantum theory. A model-independent answer was given long ago by Hui~\cite{Hui:2001wy}, who used partial-wave unitarity together with $s$-wave dominance to show that efficient elastic scattering in the collisional dark matter scenario implies $M\lesssim12\,\GeV$. (This is distinguished from the much weaker thermal relic unitarity bound, which relies on early-universe annihilation rather than present-day self-scattering~\cite{Griest:1989wd}.) The appeal of this result is its minimality: it follows from S-matrix unitarity and the nonrelativistic nature of dark matter scattering, which makes the $s$ wave dominant at low velocity. It therefore provides a robust baseline, but not necessarily the strongest constraint. As we will see, additional first-principles input, such as analyticity, crossing symmetry, locality, and gapped perturbativity, can further restrict the allowed scattering amplitude.

The modern S-matrix bootstrap has developed precisely the tools needed to sharpen such model-independent statements \cite{deRham:2025vaq, Paulos:2017fhb, Guerrieri:2021tak} (see also, e.g., \cite{Adams:2006sv, deRham:2017avq, Guerrieri:2018uew, Karateev:2019ymz, Bellazzini:2020cot, Tolley:2020gtv, Arkani-Hamed:2020blm,  Zhang:2018shp, Remmen:2020vts, Zhang:2020jyn, Gu:2020ldn, Li:2021lpe, Davighi:2021osh, Alberte:2021dnj, Li:2022rag, EliasMiro:2022xaa, Chen:2023bhu, He:2023lyy, Acanfora:2023axz,  Eckner:2024ggx, Guerrieri:2024jkn, Correia:2025uvc, Dong:2025dpy, Chang:2026ztn, Cheung:2026lpv} and \cite{deRham:2022hpx, Kruczenski:2022lot} for a review).
In this Letter, we apply these tools to derive first-principles, model-independent mass bounds for generic weakly coupled SIDM scenarios. For concreteness, we consider a scalar dark matter particle $\phi$ of mass $M$ whose interactions are weakly coupled below an energy scale $\Ls$, taken to be the cutoff of the low-energy EFT. Above it, there may be arbitrary new degrees of freedom, or the theory becomes strongly coupled. Our bootstrap approach is agnostic about the details of the high-energy/UV theory. Instead, it is assumed to satisfy a fixed-$t$ dispersion relation, which encodes analyticity, crossing symmetry and the Froissart-Martin bound of the scattering amplitudes. Partial-wave unitarity is then applied to restrict allowed 2-to-2 scattering amplitudes, which in turn constrains the SIDM parameters.

Beyond analyticity and locality, the key additional physical input compared with Hui's mass bound is the existence of a weakly coupled regime at low energies, as is commonly assumed in phenomenological model building. This assumption fixes the onset of the dominant absorptive spectrum of the S-matrix: unresolved heavy modes or loop effects are negligible below $\Ls$ at the accuracy relevant here. S-matrix consistency then places an upper envelope on the threshold amplitude $\M(4M^2,0)$ attainable by an arbitrary UV completion, which can be turned into a mass upper bound by requiring efficient self-interactions for halo phenomenology.

For a generic scalar SIDM, this yields a much stronger mass bound: $M\lesssim0.3\,\GeV$ for the benchmark $\sigma_{\rm self}=10^{-24}(M/\GeV)\cm^2$. For a pseudo-Nambu-Goldstone boson (pNGB) where the leading self-interaction is derivative dominated, the absence, or symmetry-breaking suppression, of the non-derivative $\phi^4$ interaction imposes an additional low-energy null constraint. This suppresses the threshold amplitude by powers of $M/\Ls$, further pushing the SIDM mass bound to the MeV scale. These bounds are strong enough to separate weakly coupled SIDM from the class of strongly coupled scenarios, where the low-energy scattering amplitude itself may saturate unitarity.

\noindent{\bf S-matrix bootstrap setup}
We first specify our S-matrix bootstrap framework. We use the recently developed primal bootstrap method based on fixed-$t$ dispersion relations \cite{deRham:2025vaq}. Unlike earlier primal implementations \cite{Paulos:2017fhb}, this method only uses rigorously established Martin analyticity, and constructs allowed amplitudes in the physical region. It is also particularly well suited to an EFT setup. 

Consider the elastic $\phi\phi\to\phi\phi$ amplitude $\M(s,t)$, where the Mandelstam variables obey $s+t+u=4M^2$. The partial-wave expansion is given by
\begin{equation}
\M(s,t)=16\pi\!\!\sum_{\ell~{\rm even}}\!(2\ell+1)a_\ell(s)P_\ell\!\left(1+\frac{2t}{s-4M^2}\right) ,
\label{eq:pw}
\end{equation}
where $P_\ell(x)$ is the Legendre polynomial. For a dark matter EFT weakly coupled below $\Ls>2M$, the dispersive integral over dominant, unresolved absorptive data effectively starts at $\mu=\Ls^2$. By choosing the crossing-symmetric subtraction point $s_0=t_0=u_0=4M^2/3$, the twice-subtracted fixed-$t$ dispersion relation can then be written as
\begin{align}
\label{eq:dispersion}
&\!\Rea \M(s,t)=g_0+{\cal P}\!\!\int_{\Ls^2}^{\infty}\!\!\dd\mu\!\!
\sum_{\ell~{\rm even}}\!\!\!16(2\ell+1)\,\Ima a_\ell(\mu)\\
&\times\Big\{
P_\ell\Big(1+\frac{2t}{\mu-4M^2}\Big)K_{s,t}^{\mu,t_0}+
P_\ell\Big(1+\frac{2t_0}{\mu-4M^2}\Big)K_{t,t_0}^{\mu,s_0}\Big\},\nonumber
\end{align}
where $g_0=\M(s_0,t_0)$, ${\cal P}$ denotes the principal value, and
\begin{equation}
K_{s,t}^{\mu,t_0}=\frac{1}{\mu-s}+\frac{1}{\mu-4M^2+s+t}-(s\to t_0).
\label{eq:kernel}
\end{equation}
The essence of the method is to construct all possible amplitudes by parametrizing the imaginary parts of the partial-wave amplitudes in the physical region on the right hand side of Eq.~\eqref{eq:dispersion}, and then computing the real parts from the dispersion relation. To parametrize $\Ima a_\ell(\mu)$, we map $\mu\in[\Ls^2,\infty)$ to a compact interval by defining $z=(\Ls^2-4M^2)/(\mu-4M^2)$, so that $z=1$ is the lower end of the dispersive integral and $z=0$ is infinity. We then expand
\begin{equation}
\Ima a_\ell(\mu)=\sum_{k=0}^{k_{\rm max}}c_{\ell k}\,
P_k\!\left(2\frac{\Ls^2-4M^2}{\mu-4M^2}-1\right),
\label{eq:imansatz}
\end{equation}
for even $\ell\leq \ell_{\rm max}$, where the cutoffs $\ell_{\rm max}$ and $k_{\rm max}$ are numerical truncation parameters. With this parametrization, the dispersive integrals can be evaluated explicitly, yielding expressions that are linear in the coefficients $c_{\ell k}$.

To obtain the real parts of $a_\ell(\mu)$ for $\mu>\Lambda^2$, we also perform a partial-wave expansion for $\Rea\M(s,t)$ on the left-hand side of Eq.~\eqref{eq:dispersion}, and discretize $s$ at a set of points $s_j\geq\Ls^2$. For each $s_j$, we discretize the dispersion relation on a fixed grid of $t$ values within $-(\Ls^2-4M^2)<t<0$. The resulting system of linear equations can be used to algebraically solve for $\Rea a_\ell(s_j)$. To see this more explicitly, note that, at fixed $s_j$, the left-hand side of Eq.~\eqref{eq:dispersion} is represented as 
\begin{align}
    \Rea\M(s_j,t_n)\!=\!16\pi \!\!\!\sum_{\ell~{\rm even}}\!\!(2\ell+1)\Rea a_\ell(s_j) P_\ell\Big(1\!+\!\frac{2t_n}{s_j\!-\!4M^2}\Big)  . \nonumber   
\end{align}
The values of $\Rea\M(s_j,t_n)$ are also obtained from the right-hand side of the dispersion relation as linear functions of $g_0$ and $c_{\ell k}$. Choosing $N_t=\ell_{\rm max}/2+1$ values of $t_n$ gives a square linear system for the even partial waves $\Rea a_0,\Rea a_2,\ldots,\Rea a_{\ell_{\rm max}}$ at that energy. 

Once both the imaginary and real parts of the partial waves are obtained, we impose the full partial-wave unitarity conditions. For each sampled $s_j$ and each even $\ell$, partial-wave unitarity requires
\begin{equation}
\begin{pmatrix}
1-\frac{1}{2}\rho(s_j)\Ima a_\ell(s_j) & \sqrt{\rho(s_j)}\,\Rea a_\ell(s_j)\\
\sqrt{\rho(s_j)}\,\Rea a_\ell(s_j) & 2\,\Ima a_\ell(s_j)
\end{pmatrix}\succeq0,
\label{eq:unitarity}
\end{equation}
where $\rho(s)=(1-4M^2/s)^{1/2}$. Note that the partial-wave amplitudes entering these unitarity constraints are linear functions of a common set of decision variables. Schematically, we have $\Ima a_\ell(s_j)=\sum_k B_{\ell j,k}c_{\ell k}$, $\Rea a_\ell(s_j)=A^{(0)}_{\ell j}g_0
+\sum_{\ell',k}A_{\ell j;\ell' k}c_{\ell' k}$, where the numerical matrices $A$ and $B$ are determined by the dispersion kernels and by the chosen $s$ and $t$ grids. Thus Eq.~\eqref{eq:unitarity} gives linear matrix inequalities in $g_0$ and $c_{\ell k}$. The same dispersive representation also makes the optimization objective linear:  any low-energy observable that depends linearly on the amplitude can be written schematically as
\begin{equation}
{\rm obj}[\M]=\beta_0\,g_0+\sum_{\ell,k}\beta_{\ell k}^{({\rm obj})}c_{\ell k},
\end{equation}
with the coefficients $\beta_{\ell k}^{({\rm obj})}$ fixed by the kernels in Eq.~\eqref{eq:dispersion}. Thus, the bootstrap becomes a standard, finite-dimensional semidefinite program (SDP) that can be handled by a high-precision solver such as SDPB~\cite{Simmons-Duffin:2015qma}.

\noindent{\bf Self-interaction cross section}
The SIDM $\phi\phi\to \phi\phi$ cross section is given by
\begin{align}
\sigma_{\rm self}=\frac12 \frac{1}{64\pi^2s}\int\dd\Omega\,|\M(s,t)|^2 ,
\end{align}
where we include the final-state symmetry factor $1/2$ for identical particle scattering.
In the cold dark matter limit, the relative velocity of the two incoming dark-matter particles is non-relativistic: $v_{\rm rel}\ll1$, $s=4M^2+O(M^2v_{\rm rel}^2)$ and $t=O(M^2v_{\rm rel}^2)$. Analyticity at threshold then gives 
\begin{equation}
\M(s,t)=\M_{\rm thr}
 +O\left(v_{\rm rel}^2\right) ,
\end{equation}
where we have defined $\M_{\rm thr}\equiv \M(4M^2,0)$ and possible suppressions by powers of $M/\Lambda$ have been omitted. Even for cluster velocities $v_{\rm rel}\sim10^3\,{\rm km}/{\rm s}$~\cite{Tulin:2017ara},
one has $v_{\rm rel}^2\sim10^{-5}$ in natural units. So in the present gapped EFT setup, the leading self-interaction is
effectively velocity independent on astrophysical scales,
\begin{equation}
\sigma_{\rm self}=\frac{|\M_{\rm thr}|^2}{128\pi M^2}+O(v_{\rm rel}^2) .
\label{eq:xsec}
\end{equation}
Then, comparing the bootstrap upper bound on $|\M_{\rm thr}|$ with the reference self-interaction cross section $\sigma_0=\kappa\,10^{-24}(M/\GeV)\cm^2$ gives the upper bound on the mass
\begin{equation}
M_{\rm max}=0.289\,\GeV\,\kappa^{-\frac13}\left(\frac{\max|\M_{\rm thr}|}{(4\pi)^2}\right)^{\frac23},
\label{eq:massmaster}
\end{equation}
where we have introduced $\kappa$ to parametrize the observational spread, expected to range from 0.1 to 10~\cite{Tulin:2017ara}. (For the velocity-independent $s$-wave amplitudes considered here, the total, transfer and viscosity cross sections differ only by order-one factors, which are absorbed into the parameter $\kappa$.) It remains to evaluate $\max|\M_{\rm thr}|/(4\pi)^2$. 

It is instructive to invert Eq.~\eqref{eq:massmaster} to make the origin of the mass bound more transparent. For the SIDM scenario to be relevant to halo cores, the required threshold amplitude is
\begin{equation}
\frac{|\M_{\rm thr}|}{(4\pi)^2}
=
\left(\frac{M}{0.289\,{\rm GeV}}\right)^{\frac32}
\kappa^{\frac12}.
\label{eq:alphareq}
\end{equation}
Thus a $1\,{\rm GeV}$ SIDM already requires
$|\M_{\rm thr}|/(4\pi)^2\simeq 6.4\,\kappa^{1/2}$, while a
$10\,{\rm GeV}$ SIDM requires
$|\M_{\rm thr}|/(4\pi)^2\simeq 2.0\times 10^2\,\kappa^{1/2}$.
The astrophysical self-interaction strength therefore rapidly pushes the
required threshold amplitude above the weak-coupling limit
$|\M_{\rm thr}|\lesssim (4\pi)^2$ as the mass is increased.

\noindent{\bf Numerical implementation}
The full partial-wave unitarity conditions are imposed at points $s_i=\Ls^2(1+\epsilon_s)+(s_{\rm max}-\Ls^2(1+\epsilon_s))(i/N_s)^2$, with $i=0,\ldots,N_s$, $\epsilon_s=10^{-4}$ and $N_s=299$. At each point, the partial waves are extracted from a common set of $t$ values, $t_j=-(\Ls^2-4M^2)+\epsilon_t+j(\Ls^2-4M^2-\epsilon_t)/(\ell_{\rm max}/2)$, with $j=0,\ldots,\ell_{\rm max}/2$ and $\epsilon_t=10^{-3}(\Ls^2-4M^2)$. Above $s_{\rm max}$, to speed up numerical convergence, we impose the linear unitarity conditions, namely positivity of the imaginary part of the partial waves, at $\hat s_i=s_{\rm max}/(1-i/N_{\rm lin})$ with $i=0,\ldots,N_{\rm lin}-1$ and $N_{\rm lin}=600$.

\begin{figure}[htbp]
\centering
\includegraphics[width=0.8\linewidth]{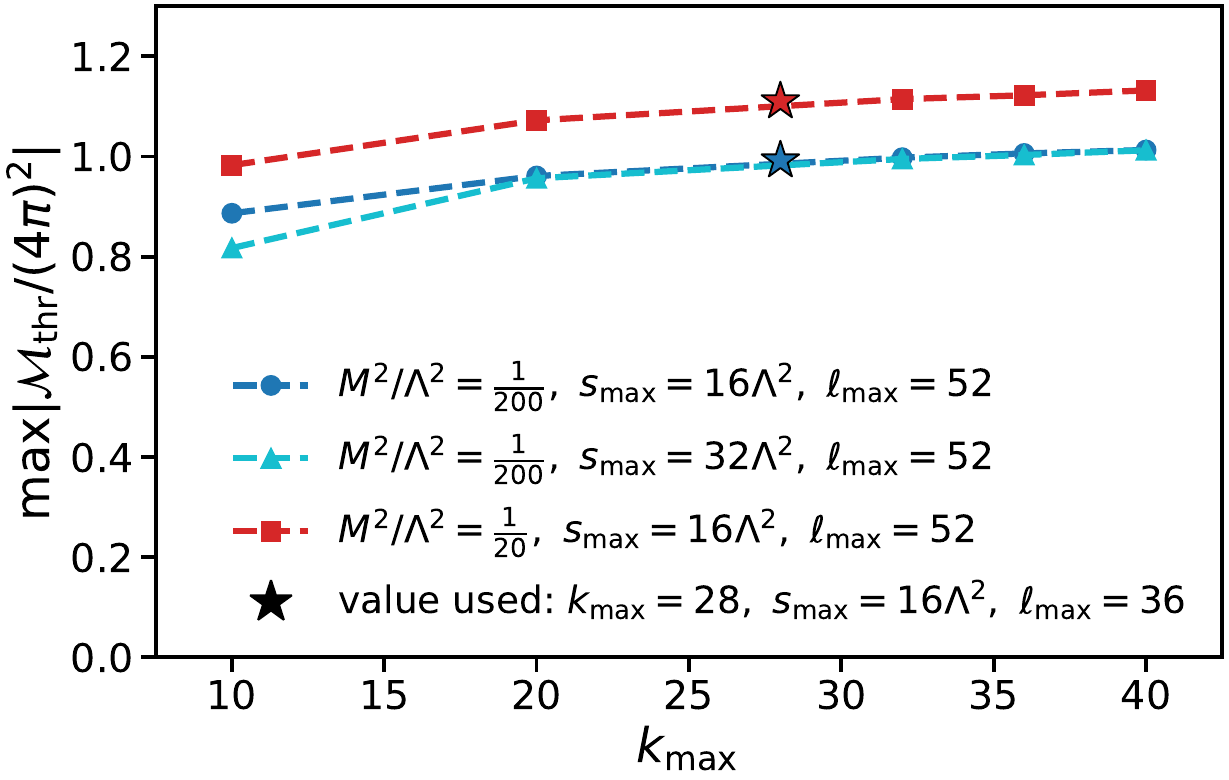}
\caption{Convergence for the threshold objective. The dashed curves show the dependence on $k_{\rm max}$ with $\ell_{\rm max}=52$, while the star markers denote the numerical parameters used in subsequent plots.}
\label{fig:conv}
\end{figure}

The convergence of the optimization objective, the maximal threshold amplitude, is shown in Fig.~\ref{fig:conv}. In the following plots we choose $s_{\rm max}=16\Ls^2$, $\ell_{\rm max}=36$ and $k_{\rm max}=28$, indicated by the star markers. Increasing $k_{\rm max}$, $\ell_{\rm max}$ or $s_{\rm max}$ changes the final dark matter mass bounds only at a level smaller than the phenomenological uncertainty in the target self-interaction cross section.

\noindent{\bf Generic weakly coupled scalar}
Let us now consider a generic scalar SIDM weakly coupled below $\Ls$. 
This is the weakly coupled counterpart of the generic collisional dark matter setup considered in Ref~\cite{Hui:2001wy}, but now we also impose analyticity, crossing symmetry and locality, in addition to partial-wave unitarity.

\begin{figure}[htbp]
\centering
\includegraphics[width=0.8\linewidth]{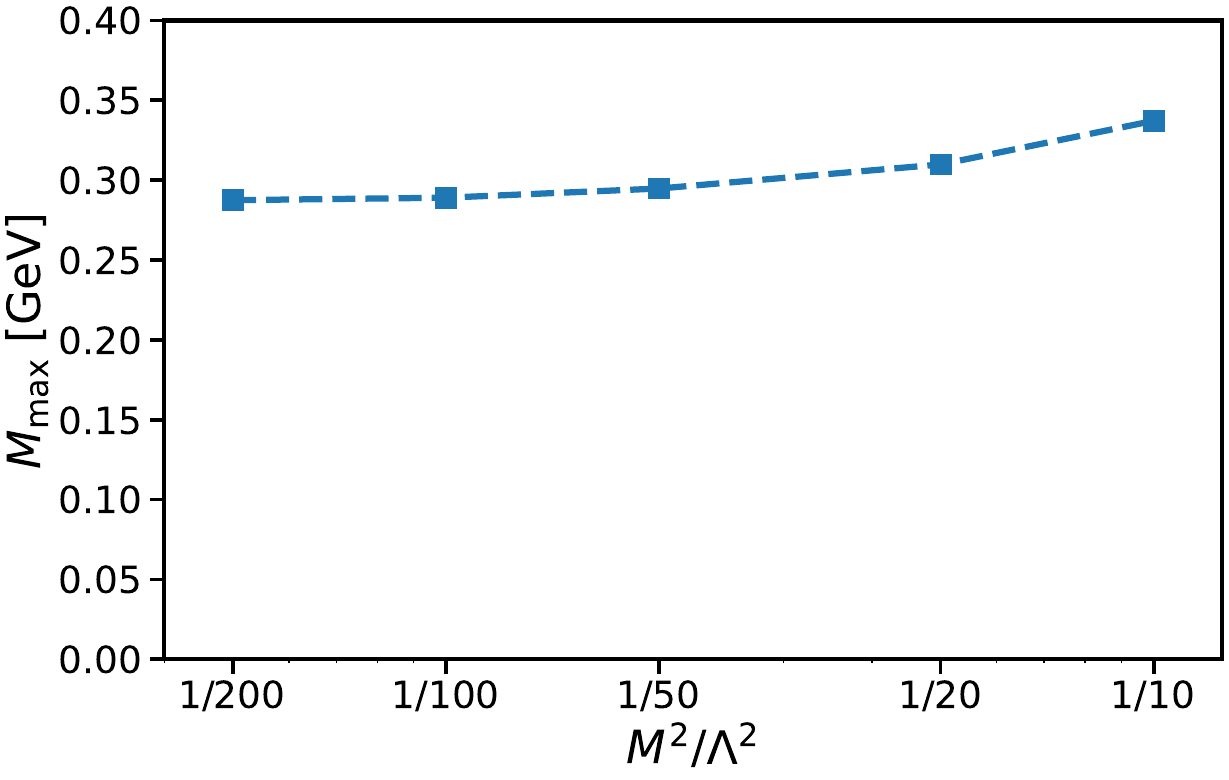}
\caption{Bootstrap bound for the dark matter mass in a generic weakly coupled SIDM scenario, for the choice of $\kappa=1$.}
\label{fig:generic}
\end{figure}

Our S-matrix bootstrap bound on the dark matter particle mass is shown in Fig.~\ref{fig:generic}. In the controlled EFT regime $M^2/\Ls^2\ll1$, the bound is nearly independent of the cutoff and gives $M_{\rm max} \simeq 0.3$\,GeV. Even when $M^2/\Ls^2$ is increased to $1/10$, close to the edge of a conservative EFT expansion, the bound grows only mildly to around $0.34$\,GeV. Using Eq.~\eqref{eq:massmaster}, we obtain the dark matter mass bound
\begin{equation}
M\lesssim 0.29\,\GeV\,\kappa^{-1/3}\quad(M^2/\Ls^2\ll1),
\label{eq:genericbound}
\end{equation}
or $M\lesssim0.34\,\GeV\,\kappa^{-1/3}$ at $M^2/\Ls^2=1/10$. Thus, the weakly coupled S-matrix bound is much stronger than the $12\,\GeV$ partial-wave unitarity bound, which allows the low-energy scattering itself to become strongly coupled. This is the first main result. Partial-wave unitarity alone allows an arbitrary threshold amplitude until a partial wave saturates, while our dispersive bootstrap correlates the threshold amplitude with the entire crossing-symmetric absorptive spectrum above the higher scale $\Ls$.

\noindent{\bf Pseudo-Nambu-Goldstone scalar}
Next, we turn to the interesting subclass of SIDM models where the dark matter scalar is a pseudo-Nambu-Goldstone boson (pNGB)~\cite{Bai:2010qg,Frigerio:2012uc,Cline:2013zca,Hochberg:2014kqa,Kribs:2016cew,Contino:2020god,Abe:2024lzj,Sheikh:2025qym}. We are particularly interested in the scenario in which the leading effective 2-to-2 scattering is dominated by the leading shift-symmetric interaction $(\partial\phi)^4$. The softness due to the derivative suppression at low energies provides an elegant mechanism to naturally evade stringent direct-detection bounds at zero-momentum transfer~\cite{Gross:2017dan, Abe:2024lzj}, while maintaining non-trivial self-scattering signatures at finite momentum transfer relevant for halos. Such a scenario is natural if the NGB shift symmetry is softly broken only by a mass term, in which case no additional shift-symmetry-breaking self-interactions are generated radiatively and the mass term affects the amplitude only through the on-shell kinematics~\cite{deRham:2017imi}. More generally, pNGBs can contain non-derivative interactions, for example, in chiral perturbation theory, or in axion-like models with periodic potentials generated by instanton or other non-perturbative effects. In such models, a Lagrangian term of the form ${\cal L}\supset M^2\phi^4/F^2$ is typically generated. If $F$ is a much higher scale ($F\gg \Lambda^2/M$), the leading 2-to-2 scattering is still  $(\partial\phi)^4$ dominated. This happens when the shift-symmetry breaking and non-derivative terms are generated at loop level~\cite{Giudice:2007fh,Panico:2015jxa,deBoer:2025oyx,Agashe:2004rs}, as in models where the dark matter also couples to the Standard Model gauge bosons or fermions \cite{Bai:2010qg,Frigerio:2011in,Contino:2020god,Cai:2026avj}.

The dominance of the $(\partial\phi)^4$ interaction, or equivalently the suppression of the $\phi^4$ term, implies an accidental null condition among the low-energy amplitude coefficients, which must be imposed in the bootstrap to extract the bound. To implement this condition, we expand the crossing-symmetric amplitude around $s_0=t_0=u_0=4M^2/3$ as 
\begin{align}
\label{eq:Mfof2}
\! \M=\!f_0+f_2[(s\!-s_0)^2\!+\!(t\!-t_0)^2\!+\!(u\!-u_0)^2
]+\cdots .
\end{align}
In this pNGB scenario, because the leading interaction is $(\partial\phi)^4$ and the $\phi^4$ interaction is absent or suppressed by $O({M^2}/{F^2})$, $f_0$ is not independent: 
\begin{equation}
f_0-\frac{4}{3}M^4 f_2=0 .
\label{eq:null}
\end{equation}
This follows by matching ${\cal M} \propto s^2+t^2+u^2-4M^4+\cdots$ to Eq.~(\ref{eq:Mfof2}) to leading order.
This null constraint implies that the leading threshold amplitude is soft: $\mathcal{M}_{\mathrm{thr}}\simeq 12 M^4 f_2=O(M^4/\Lambda^4)$. We impose Eq.~\eqref{eq:null} as a null constraint in the SDP. Corrections from higher-derivative operators or additional soft-breaking interactions are suppressed by further powers of $M^2/\Ls^2$ and can be included once a specific pseudo-Goldstone model is specified.

\begin{figure}[htbp]
\centering
\includegraphics[width=0.8\linewidth]{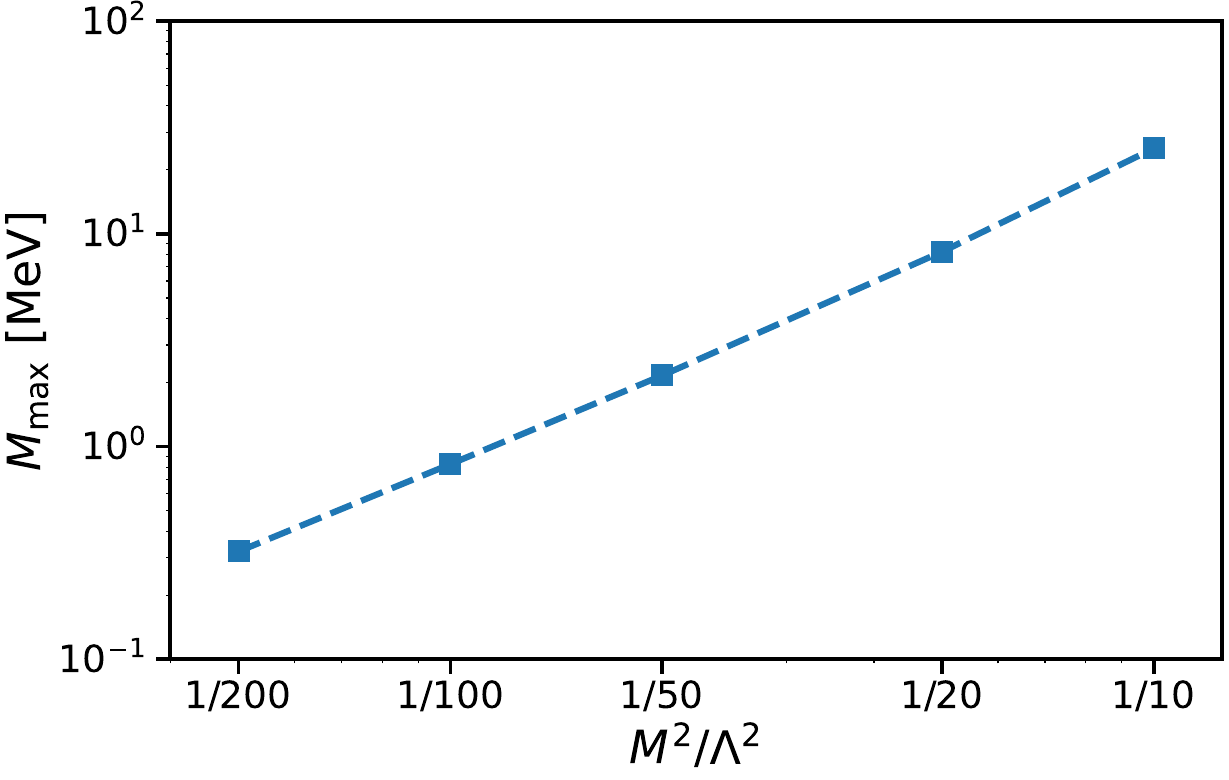}
\caption{Bootstrap bound for the dark matter mass in pNGB SIDM scenario, which is suppressed by powers of $M/\Ls$, for the choice of $\kappa=1$.}
\label{fig:pNGB}
\end{figure}

The resulting bound is shown in Fig.~\ref{fig:pNGB}. The scaling is now changed qualitatively: the threshold amplitude is forced to vanish in the $M/\Ls\to0$ limit. The numerical bootstrap makes this suppression quantitative. The upper envelope is well described by
\begin{equation}
\frac{\max|\M_{\rm thr}|}{(4\pi)^2}
\simeq C_{\rm pNGB}
\left(\frac{M}{\Ls}\right)^4,
\label{eq:pNGBscaling}
\end{equation}
where $C_{\rm pNGB}=O(1)$. Equivalently, for a fixed hierarchy $\epsilon\equiv M/\Ls$, Eq.~\eqref{eq:massmaster} gives
\begin{equation}
M_{\rm max}
\simeq
0.289\,{\rm GeV}\,
C_{\rm pNGB}^{\frac23}
\kappa^{-\frac13} \epsilon^{\frac83} .
\label{eq:pNGBetabound}
\end{equation}
Thus, in the pNGB case, the SIDM mass bound is parametrically strengthened by the separation between the dark matter mass and the EFT cutoff. A relatively mild separation $M^2/\Ls^2=1/10$ gives approximately
\begin{equation}
M\lesssim 26\,{\rm MeV}\,\kappa^{-1/3},
\label{eq:pNGBbound}
\end{equation}
while a more controlled EFT with $M^2/\Ls^2\leq1/100$ pushes the bound to the MeV or sub-MeV range. In other words, because derivative interactions are soft at threshold, pseudo-Goldstone SIDM that remains weakly coupled up to a separated cutoff is constrained to be much lighter.

\noindent{\bf Summary and discussion}
We have derived first-principles S-matrix-bootstrap bounds on the mass of weakly coupled scalar self-interacting dark matter. The basic logic is that halo phenomenology fixes the size of a low-energy scattering amplitude, while analyticity, crossing symmetry, locality and partial-wave unitarity bound how large this amplitude can be if the dark matter remains weakly coupled below a separated scale $\Ls$. Rather than assuming a particular UV completion, we parametrize the absorptive partial waves above $\Ls$ and reconstruct the amplitude through a fixed-$t$ dispersion relation. The resulting optimization scans over arbitrary heavy states or strong dynamics above the cutoff, subject only to S-matrix consistency.

For the benchmark SIDM cross section
$\sigma_{\rm self}=10^{-24}(M/\GeV)\cm^2$, we find that the mass of a generic weakly coupled SIDM particle satisfies $M\lesssim 0.3\,\GeV$.
This is stronger than the classic partial-wave unitarity bound $M\lesssim12\,\GeV$, since the latter permits strong coupling near threshold. For a pseudo-Nambu-Goldstone scalar, the absence or suppression of the non-derivative $\phi^4$ interaction imposes an additional null constraint on the crossing-symmetric amplitude. This makes the threshold amplitude soft in the $M/\Ls\to0$ limit and pushes the allowed mass to $M\lesssim O(\text{MeV})$. The bounds scale with the observational target as $M_{\rm max}\propto \kappa^{-1/3}$, so the conclusion is only mildly affected by the precise value of $\sigma_{\rm self}/M$ within the usual SIDM range.

The strength of the bound has a simple physical origin. At fixed $\sigma_{\rm self}/M$, increasing the dark matter mass requires a rapidly growing threshold amplitude. The bootstrap asks whether the required infrared amplitude can be embedded into a crossing-symmetric, analytic and unitary S-matrix whose unknown absorptive spectrum starts only at $\Ls$. Once this is imposed, the threshold amplitude cannot be adjusted independently of the high-energy absorptive data. Thus the result is not merely a perturbativity estimate; it excludes would-be weakly coupled EFTs whose required threshold scattering is too large to admit a consistent completion with the assumed gap.

The logic of our dispersive bound is to use UV unitarity, encoded in a
UV-agnostic parametrization of the absorptive data above $\Lambda$, to constrain the IR amplitude \cite{deRham:2022hpx}. This is different from the commonly used perturbative unitarity bound, which instead applies unitarity directly within the EFT below $\Lambda$. More importantly, in the present nonrelativistic SIDM context, the perturbative unitarity estimate is parametrically of the same order as the nonperturbative bound of Ref.~\cite{Hui:2001wy}. The reason is that, near threshold, the phase-space factor in partial-wave unitarity, $\rho(s)\sim v_{\rm rel}/2$, vanishes, allowing a saturated partial-wave amplitude to scale as $1/\rho\sim 1/v_{\rm rel}$. This gives the familiar estimate $\sigma_{\rm self}\lesssim O((Mv_{\rm rel})^{-2})$, corresponding to a strongly coupled threshold amplitude.

Let us also clarify the scope of the assumptions. The UV completion above $\Ls$ need not be perturbative, narrow-resonance dominated, or described by a small number of particles. All such possibilities are included in the absorptive partial waves scanned over by the bootstrap. The assumption is instead infrared: below $\Ls$, the dark matter particle is a weakly coupled scalar EFT degree of freedom.

Several extensions are immediate. The same dispersive bootstrap can be applied to dark matter with spin or internal symmetry, and to multi-channel systems. Although technically more involved, one may expect them to lead to comparable, or even stronger, mass bounds. It can also incorporate additional low-energy information, such as approximate chiral symmetry and dark gauge symmetry.

~\\

\section*{Acknowledgments}
We would like to thank Fu-Ming Chang and Zhuo-Yan Chen for helpful discussions. SYZ acknowledges support from the National Natural Science Foundation of China under grant No.~12475074 and No.~12247103. QC is supported by the National Natural Science Foundation of China under Grant No.~12505118, Anhui Provincial Natural Science Foundation under Grant No.~2508085QA034, and the Scientific Research Foundation for High-level Talents of Anhui University of Science and Technology under Grant No.~2024yjrc166. This research is also supported by the advanced computing resources provided by the Supercomputing Center of the USTC.

\bibliography{dm_bootstrap_bounds}

\end{document}